\documentclass[
aps,prd,twocolumn,reprint,amsmath,amssymb,preprintnumbers,superscriptaddress,nobibnotes,nofootinbib,floatfix,longbib
]{revtex4-1}

\usepackage{graphicx}
\usepackage{dcolumn}
\usepackage{bm}
\usepackage{color}
\usepackage{amssymb}
\usepackage{amsmath}
\usepackage{braket}
\usepackage[colorlinks=true,allcolors=Blue,pdfusetitle]{hyperref}
\usepackage{rotating}
\usepackage{multirow}
\usepackage{graphicx}
\usepackage[dvipsnames]{xcolor}
\usepackage{svg}
\usepackage{esdiff}
\definecolor{Green}{RGB}{0, 128, 0}
\newcommand{\orcid}[1]{\href{https://orcid.org/#1}{\includegraphics[width=10pt]{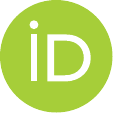}}}

\begin{document}
\preprint{LLNL-JRNL-2000909, FERMILAB-PUB-24-0810-T, N3AS-24-037}
\title{Exploring entanglement and spectral split correlations in three-flavor collective neutrino oscillations}

\author{Pooja Siwach \orcid{0000-0001-6186-0555}}
\email{siwach1@llnl.gov}
\affiliation{Nuclear and Chemical Science Division, Lawrence Livermore National Laboratory, Livermore, California 94551, USA}

\author{A. Baha Balantekin \orcid{0000-0002-2999-0111}}
\email{baha@physics.wisc.edu}
\affiliation{
Department of Physics, University of Wisconsin--Madison,
Madison, Wisconsin 53706, USA}

\author{Amol V. Patwardhan \orcid{0000-0002-2281-799X}}
\email{apatwa02@nyit.edu}
\affiliation{Department of Physics, New York Institute of Technology, New York, NY, 10023}
\affiliation{School of Physics and Astronomy, University of Minnesota, Minneapolis, MN 55455}

\author{Anna M. Suliga \orcid{0000-0002-8354-012X}}
\email{a.suliga@nyu.edu}
\affiliation{Center for Cosmology and Particle Physics, New York University, New York, NY 10003, USA}
\affiliation{
Department of Physics, University of California Berkeley, Berkeley, California 94720, USA}
\affiliation{
Department of Physics, University of California, San Diego, La Jolla, CA 92093-0319, USA}


\date{\today}


\begin{abstract}
In environments with prodigious numbers of neutrinos, such as core-collapse supernovae, neutron star mergers, or the early universe, neutrino-neutrino interactions are dynamically significant. They can dominate neutrino flavor evolution and force it to be nonlinear, causing collective neutrino oscillations. Such collective oscillations have been studied numerically, for systems with up to millions of neutrinos, using mean-field or one-particle effective approximations. However, such a system of interacting neutrinos is a quantum many-body system, wherein quantum correlations could play a significant in the flavor evolution--thereby motivating the exploration of many-body treatments which follow the time evolution of these correlations. In many-body flavor evolution calculations with two neutrino flavors, the emergence of spectral splits in the neutrino energy distributions has been found to be correlated with the degree of quantum entanglement across the spectrum. In this work, for the first time, we investigate the emergence of spectral-splits in the three-flavor many-body collective neutrino oscillations. We find that the emergence of spectral splits resembles the number and location found in the mean-field approximation but not in the width. Moreover, unlike in the two-flavor many-body calculations, we find that additional degrees of freedom make it more difficult to establish a correlation between the location of the spectral splits and the degree of quantum entanglement across the neutrino energy spectrum.
\end{abstract}

\maketitle


\section{Introduction}
\label{sec:introduction}

Neutrino flavor evolution in environments with extreme number of these particles is affected not only by their interactions with matter~\cite{Wolfenstein:1979ni, Mikheev:1986if} but also themselves~\cite{Fuller:1987gzx,  Notzold:1987ik, Pantaleone:1992eq, Pantaleone:1992xh, Samuel:1993, Sigl:1993ctk, Qian:1994wh, Kostelecky:1995} leading to collective neutrino oscillations~\cite{Duan:2009cd, Duan:2010bg, Chakraborty:2016yeg, Tamborra:2020cul, Richers:2022zug, Volpe:2023met}.
This can affect the transport of the energy, entropy, and lepton number due to the tight coupling of electron neutrinos and antineutrinos with matter established by $\nu_e$ charged-current interactions with free nucleons and heavy nuclei. Therefore, it is crucial to understand the neutrino flavor evolution in these environments, including core-collapse supernovae, binary neutron star mergers, neutron star-black hole mergers, and specific periods of the early universe~\cite{Janka:2006fh, Burrows:2020qrp, Fuller:2022nbn, Foucart:2022bth, Kyutoku:2017voj, Grohs:2015tfy}. In addition, the collective neutrino oscillations can also affect neutrino decoupling from matter~\cite{Raffelt:2001kv, 2012PhRvD..85h3003F, Shalgar:2022rjj, Shalgar:2022lvv}, which influences the efficiency of neutrino heating, and hence, could impact the core-collapse supernova explosion as well as nucleosynthesis~\cite{Surman:2003qt, Steigman:2012ve, Martinez-Pinedo:2017ksl, Langanke:2019ggn, Kajino:2012zz, Frohlich:2015spx, Roberts:2016igt, Grohs:2015tfy, Xiong:2019nvw, Xiong:2020ntn, Wu:2014kaa, Qian:1993dg, Kajino:2014bra, Yoshida:2006qz, Balantekin:2017bau, Balantekin:2018mpq, Duan:2010af, Sasaki:2017jry, George:2020veu, Balantekin:2023ayx}.

Due to the sheer number of neutrinos interacting among themselves in these environments triggering many-body effects, some approximations must be used to simulate the collective neutrino oscillations. One of the most fundamental is the mean-field approximation~\cite{Samuel:1993, Sigl:1993ctk, Qian:1994wh, Balantekin:2006tg, Pehlivan:2011hp}. It simplifies the problem by assuming that the flavor evolution can be described by looking at a single neutrino interacting by one-body term interaction with a background of all other neutrinos. Such a description may not be able to completely capture or discern the effects of quantum correlations appearing in the many-body calculations~\cite{Bell:2003mg, Friedland:2003dv, Friedland:2003eh, Friedland:2006ke, McKellar:2009py, Pehlivan:2011hp, Pehlivan:2014zua, Birol:2018qhx, cervia:2019, Roggero:2021asb, Patwardhan:2021rej, Cervia:2022pro, Lacroix:2022krq, Patwardhan:2022mxg, Balantekin:2023qvm, Xiong:2021evk, Martin:2021bri, Martin:2023ljq, Roggero:2021, Roggero:2022hpy, Illa:2023prl, Rrapaj:2019pxz, Martin:2023gbo, Bhaskar:2023sta, Lacroix:2024pbb, Neill:2024klc, Cirigliano:2024pnm}. The treatment of the neutrino flavor evolution in the many-body framework has also been found to possibly change the nucleosynthesis yields in supernova environments as compared to the corresponding mean-field oscillation calculations~\cite{Balantekin:2023ayx}. Some studies have also expressed caution regarding the use of interacting plane waves or neutrino beams in many-body neutrino oscillation treatments, and have proposed modifications to this approach, such as incorporating finite interaction length/time~\cite{Shalgar:2023ooi, Kost:2024esc}, or momentum-changing (non-forward) neutrino scattering~\cite{Johns:2023ewj,Cirigliano:2024pnm}. An alternative framework for describing quantum correlations in collective oscillations is based on the Bogoliubov-Born-Green-Kirkwood-Yvon (BBGKY) hierarchy~\cite{Volpe:2013uxl}.

The quantification of quantum correlations in the collective neutrino oscillations in terms of the quantum entanglement provides useful insights into the nature of these correlations and hence the complexity of the problem \cite{cervia:2019, Roggero:2021asb, Patwardhan:2021rej, Cervia:2022pro, Lacroix:2022krq, Illa:2023prl}. Therefore, collective neutrino oscillations are an interesting playground to explore the quantum information science tools.
For instance, the many-body studies under two-flavor approximation, using the single-angle approximation, have found that the entanglement entropy is largest for neutrino energy modes closest to the spectral splits~\cite{Patwardhan:2021rej}. Spectral splits are a distinct feature appearing in the collective neutrino oscillations, where neutrinos of different flavors fully or partially exchange their energy spectra~\cite{Birol:2018qhx, Duan:2006an, Duan:2006prl, Raffelt:2007prd, Raffelt:2007prd1, Duan:prd2007, Duan:prd20071}. While this phenomenon was initially discovered using two-flavor mean-field calculations, subsequent three-flavor calculations were found to exhibit \textit{multiple} spectral splits among different flavor states~\cite{Fogli:2008pt, Fogli:2008fj, Dasgupta:2009mg, Fogli:2009rd, Dasgupta:2010cd}. 

In the many-body calculations, the survival probabilities of neutrinos with frequencies closer to the spectral split frequency were found to deviate the most from the mean-field calculations~\cite{cervia:2019, Patwardhan:2021rej}. Based on these observations, it was suggested to adopt a hybrid approach by performing many-body simulations for neutrinos closer to the spectral splits and mean-field approximation for the remaining neutrinos. In addition, the spectral splits in the many-body settings have been found to be much broader than in the mean-field calculations and the width of the spectral split seems to be correlated with the mixing angle~\cite{Patwardhan:2021rej}. We aim to extend these findings to the three-flavor problem under many-body effects, using the single-angle approximation in this study. A recent many-body calculation using two flavors has shown that multi-angle effects may inhibit the development of spectral splits for certain neutrino configurations~\cite{Martin:2023ljq}. However, we shall defer the multi-angle analysis of the three-flavor system to future work.

We calculated the collective neutrino oscillations in many-body picture for three-flavor case for the first time in Ref.~\cite{siwach:2023prd}, but limited to an ensemble of five interacting neutrinos. We demonstrated that the two-flavor many-body calculations can underestimate the entanglement between neutrinos compared to the three-flavor case. We have also found that a new properties emerge for quantum systems of dimensions three or higher~\cite{Balantekin:2024pwc}.

Here, we extend three-flavor many-body calculations to seven neutrinos and demonstrate the scaling of entanglement entropy with the number of neutrinos. We further investigate the emergence of spectral splits and the correlation between the entanglement entropy and the spectral split frequency. In addition, we compare our findings to mean-field calculations. To explore the three-flavor systems in the many-body picture for a larger ensemble require the employment of quantum computers. Such efforts are already made for the two-flavor case~\cite{Hall:2021rbv, Yeter-Aydeniz:2021olz, Amitrano:2022yyn,Illa:pra2022, Siwach:2023wzy} as well as three-flavor case~\cite{Chernyshev:2024kpu, Turro:arxiv2024}, but the hardware implementations are all qubit-based, even for three flavors. Simulating the three-flavor case with qutrits are more natural way due to a direct mapping between three levels of qutrits and three flavors of a neutrino.



 
This paper is organized as follows. In Sec.~\ref{sec:Hamiltonian}, we briefly review the formalism used to treat the three-flavor collective neutrino oscillations in the many-body picture and Sec.~\ref{sec:Mean-Field} in the mean-field approximation. 
We present the results for the spectral splits in the mean-field and many-body calculations in Sec.~\ref{sec:Results}. We conclude and discuss the main differences and similarities in the emergence of spectral splits in the two approaches in Sec.~\ref{sec:Conclusions}.
 
\section{Formalism}
\label{sec:Hamiltonian}
The time-evolution of neutrino flavor exhibiting collective oscillations can be minimally described by a Hamiltonian comprising the vacuum propagation and neutrino-neutrino interaction terms~\cite{Notzold:1987ik, Sigl:1993ctk}. We ignore the matter effects~\cite{Wolfenstein:1977ue, Mikheev:1986if} assuming that these effects may not make a qualitative difference to the outcome in the regime where neutrino-neutrino interaction is dominant~\cite{Pehlivan:2011hp, Pehlivan:2014zua, Birol:2018qhx,Cervia:2019nzy, Patwardhan:2019zta, Rrapaj:2019pxz, Patwardhan:2021rej, Cervia:2022pro, Lacroix:2022krq}. Therefore, we have 
\begin{equation}
\label{eq:H}
    H = H_{v}+H_{\nu\nu} \ ,
\end{equation}
where $H_{v}$ and $H_{\nu\nu}$ account for the vacuum oscillations (one-body) and neutrino-neutrino interaction (two-body) term, respectively. The exact form of the Hamiltonian in SU(3) generator representation can be written as~\cite{Pehlivan:2014zua,siwach:2023prd}
\begin{equation}
\label{eq:H_Q}
    H=\sum_p \vec{B}\cdot\vec{Q}_p+\sum_{p,p'}\mu_{pp'}\vec{Q}_p\cdot\vec{Q}_{p'} \ ,
\end{equation}
where $\vec{B}$ in the mass basis is given by
\begin{equation}
\label{eq:B}
    \vec{B}=\left(0,0,\omega_p,0,0,0,0,\frac{2}{\sqrt{3}}\Omega_p\right) \ ,
\end{equation}
with the neutrino frequencies $\omega_p=-{\delta m^2}/{2E}$ and $\Omega_p=-{\Delta m^2}/{2E}$. $E$ is the energy of the neutrino in $p^{\rm th}$ mode, the neutrino squared mass differences are $\delta m^2=m_2^2-m_1^2$ and $\Delta m^2\approx|m_3^2-m_2^2|\approx|m_3^2-m_1^2|$. The sign of the larger squared mass difference $\Delta m^2$ is not fixed and hence, in this work, we consider both mass orderings called normal (NO) and inverted (IO) mass orderings. The generators $Q_p$ can be expressed as
\begin{equation}
\label{eq:Q}
    Q_{i'} = \frac{1}{2}\sum_{i,j=1}^{3}a_i^{\dagger}(\lambda_{i'})_{ij}a_j \ ,
\end{equation}
with $\lambda$'s as the $3\times3$ Gell-Mann matrices ($i' \in \{1, \ldots,8 \}$) and $a_i(a_i^{\dagger})$ are the fermionic annihilation (creation) operators.

The neutrino-neutrino interaction strength is given by
\begin{equation}
\label{eq:mu_pp}
    \mu_{pp'} = \frac{\sqrt 2 G_F}{V}(1-\cos\theta_{pp'}) \ ,
\end{equation}
where $V$ and $G_F$ are the quantization volume and Fermi constant, respectively. $\theta_{pp'}$ is the angle between the trajectories of two neutrinos with momenta $p$ and $p'$. To further simplify the simulations, we use the ``single-angle" approximation~\cite{Duan:2006an,Qian:1994wh, Bell:2003mg, Friedland:2006ke} in a neutrino bulb geometry, under which the strength parameter is given by
\begin{equation}
\label{eq:mu_r}
    \mu(r)=\frac{G_F}{\sqrt2 V}\left(1-\sqrt{1-\frac{R_{\nu}^2}{r^2}}\right)^2 \ ,
\end{equation}
where $R_{\nu}$ is the radius of the neutrinosphere and $r$ is the distance from the center of the supernova.

We solve the time-dependent Schr\"odinger equation to simulate the flavor evolution under Hamiltonian given in Eq.~\eqref{eq:H_Q}. The evolution of entanglement can be understood by calculating the one-body von-Neumann entropies for each neutrino, given by
\begin{equation}
    S_n = -{\rm Tr}[\rho_n\log\rho_n] \ ,
\end{equation}
where $\rho_n$ is the reduced density matrix for the $n$th neutrino. Another important quantity we calculate is the polarization vector $(\vec{P})$ which provides a direct insight into the spectral splits. The polarization vector can be calculated component-wise as
\begin{equation}
    P_{nj}={\rm Tr}[\rho_n\lambda_j] \ ,
\end{equation}
where
\begin{equation}\label{eq:sp_density}
    \rho_n=\frac{1}{3}\left[I+\frac{3}{2}\sum_{j=1}^8\lambda_jP_{nj}\right] \ .
\end{equation}
Unlike the two-flavor case where only one component of the polarization vector is conserved, in the three-flavor case two such components $P_3$ and $P_8$ are conserved, making the analysis more complex. As shown in Ref.~\cite{siwach:2023prd}, there is a direct correlation between the entanglement and the polarization vector.


To better interpret the spectral-splits in many-body calculations, first we perform the mean-field calculations and interpret the location, width and number of spectral splits within that approximate treatment. 

\subsection{Mean-Field}
\label{sec:Mean-Field}

In the mean-field limit, neutrinos are no longer entangled; instead they all interact with the same mean field. Hence, in that limit the Hamiltonian which describes the neutrinos becomes a one-body Hamiltonian. After also averaging the $\mu_{p,p'}$ term over all the neutrino momenta (single-angle approximation), the mean-field limit involves replacing the second term in Eq.~\eqref{eq:H_Q} by
\begin{equation}
   \sum_{p,p'}\mu_{pp'}\Vec{Q_{p}}\cdot\Vec{Q_{p'}} \rightarrow \mu (r) \sum_p\Vec{Q_{p}}\cdot\sum_{p'}\langle\Vec{Q_{p'}} \rangle  \ ,
\end{equation}
where $\langle .. \rangle$ represents an average over a suitably defined state. Hence Eq. (\ref{eq:H_Q}) takes the form 
\begin{equation}
    \label{mf_H}
    H_\mathrm{MF} = \sum_{p}\Vec{B}\cdot\Vec{Q_{p}}+ 
    \mu (r) \sum_p\Vec{Q_{p}}\cdot\sum_{p'}\langle\Vec{Q_{p'}} \rangle \ . 
\end{equation} 
In Ref.~\cite{Balantekin:2006tg}, it was explicitly shown that, if the states used for averaging are the SU(2) coherent states, then the mean-field corresponds to the saddle point approximation of the path integral representing the evolution of the many-neutrino system.

Since neutrinos are no longer entangled in the mean-field limit the density matrix of each neutrino satisfies the equation
\begin{equation}
\label{denmat}
    i \frac{\partial}{\partial t} \rho (p) = [ H_\mathrm{MF}, \rho (p) ] \ . 
\end{equation}
Substituting the single-particle density matrix $\rho (p)$ given in Eq.~\eqref{eq:sp_density}
into the Eq. (\ref{denmat}) one obtains 
\begin{equation}
\frac{\partial P_a(p)}{\partial t}= f_{abc}
\left(B_b(\omega_{p},\Omega_{p})+\mu(r)\Pi_b\right)P_c (p) \ ,
\end{equation}
where we defined $\vec{\Pi} = \sum_p \vec{P} (p)$. 
Using generalized vector product, {\it i.e.,} writing $A_a= f_{abc} B_b C_c$ as $\vec{A} = \vec{B} \times \vec{C}$, where $f_{abc}$ are the structure coefficients, we can rewrite this equation as 
\begin{eqnarray}\label{eq:mf_final}
    \frac{\partial \vec{P}(p)}{\partial t}=\left(\vec{B}(\omega_{p},\Omega_{p})+\mu(r)\vec{\Pi}\right)\times \vec{P}(p) \ .
\end{eqnarray}

The detailed equations are given in Appendix~\ref{sec:mf_Eq}.
Solving the above equation, we obtain the trajectory of the polarization vector for each neutrino in the mean-field.
\section{Results}
\label{sec:Results}
\begin{figure}
    \centering
    \includegraphics[width=0.95\linewidth]{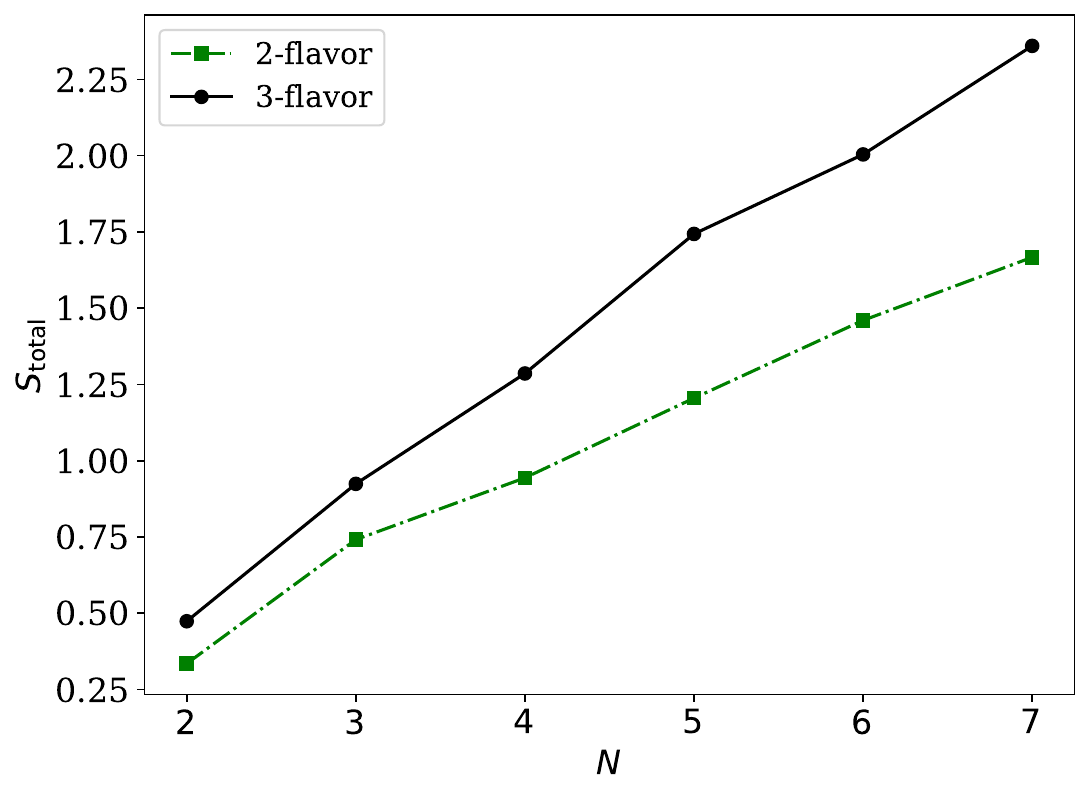}
    \caption{The total entanglement entropy as a function of number of neutrinos in the system $N$ in two-flavor and three-flavor settings for an initial state with all neutrinos in electron flavor $\nu_e$.}
    \label{fig:total-S}
\end{figure}

\begin{figure*}[t]
    \centering
    \includegraphics[width=0.3\linewidth]{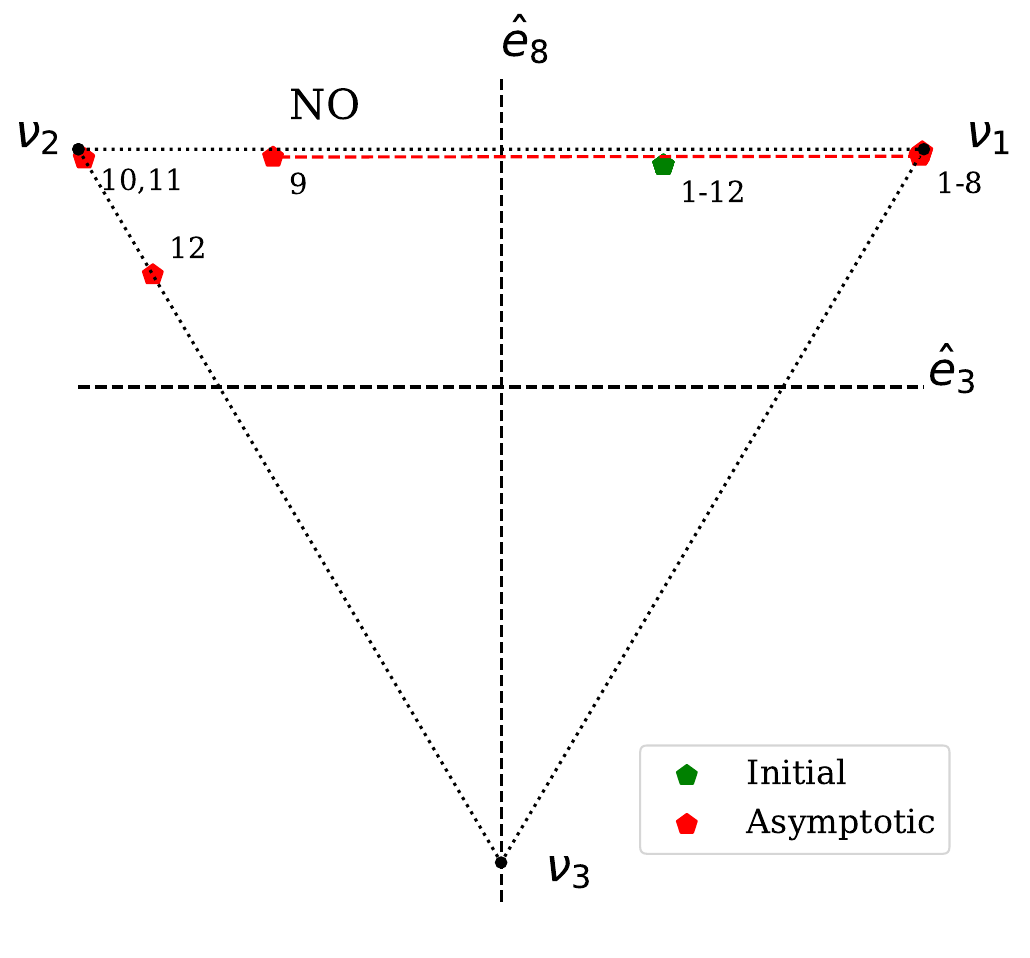}
    \includegraphics[width=0.3\linewidth]{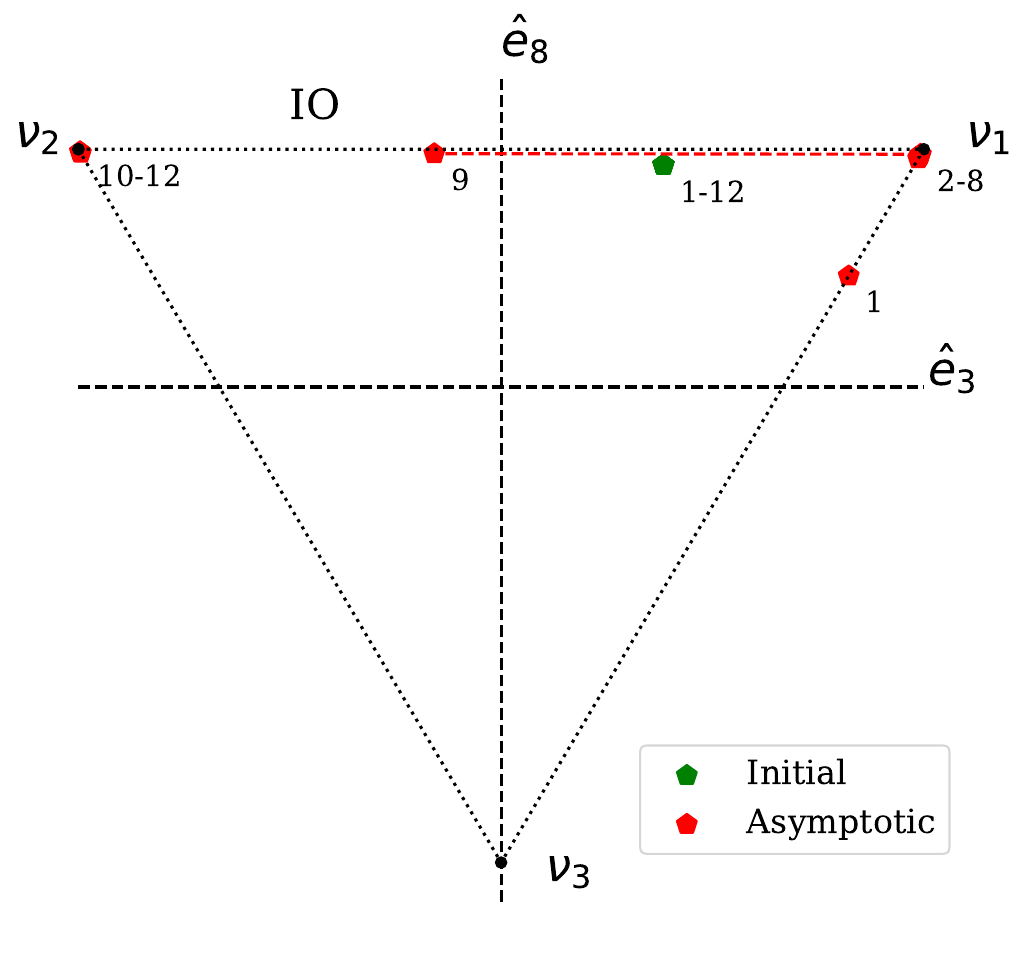}
    \includegraphics[width=0.3\linewidth]{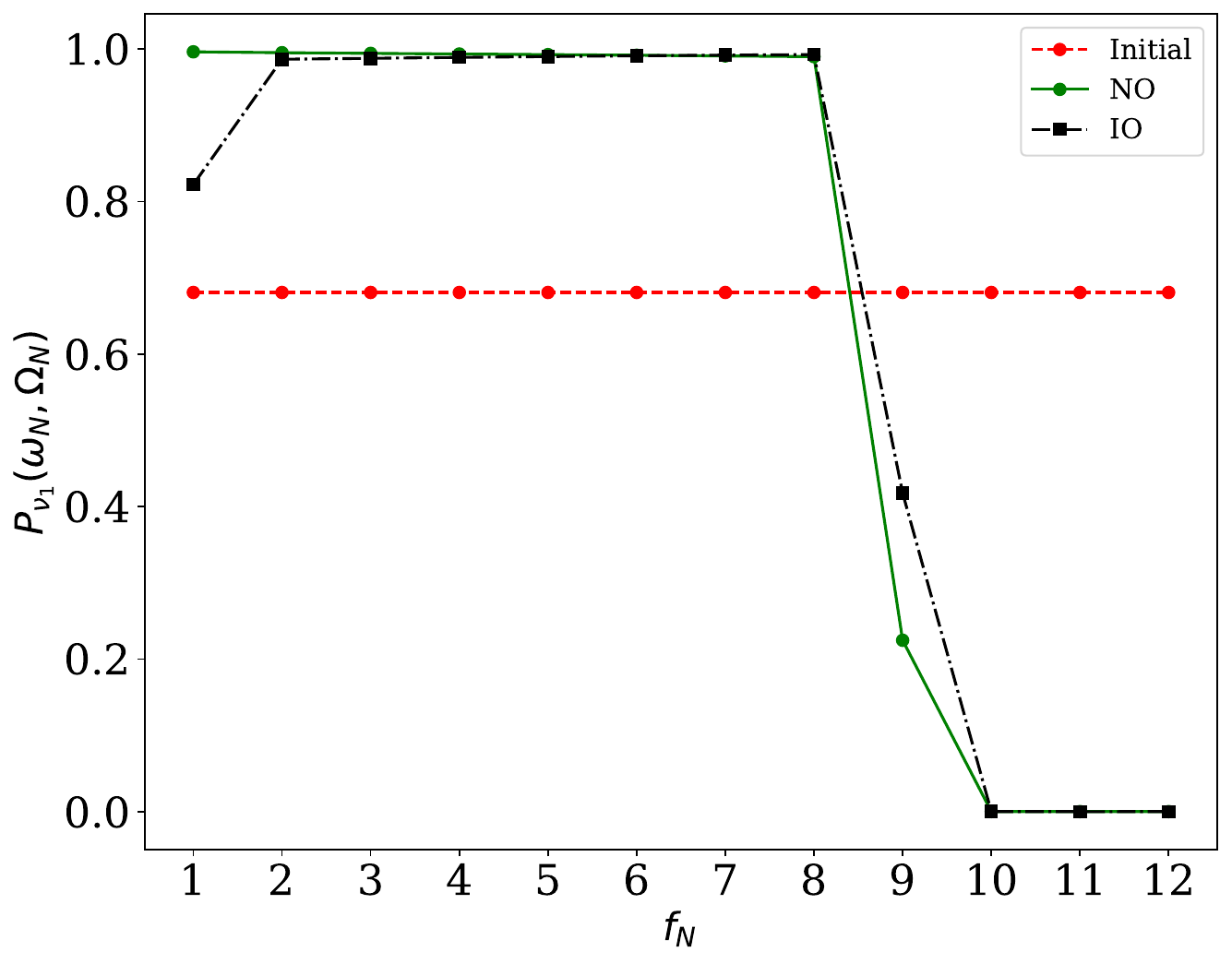}
    \caption{The asymptotic $P_3$ and $P_8$ values in the $\hat{e}_3-\hat{e}_8$ plane for $N=12$ neutrinos within the mean-field in the NO (\emph{left panel}) and the IO (\emph{right panel}) cases with initial state $\ket{\nu_e^{\otimes 12}}$. The points are labeled by the neutrino frequency mode, i.e., q, and the spectral splits are shown by red dotted lines. \emph{Right panel:} The initial (red circles) and asymptotic values of $P_{\nu_1}$ for NO (green circles) and IO (black squares) as a function of the neutrino frequency mode number $f_N$.}
    \label{fig:N12_e}
\end{figure*}
We extend the many-body simulations of three-flavor collective neutrino oscillations up to an ensemble of seven neutrinos as compared to five neutrinos in our previous work. The total entanglement entropy, {\it i.e.}, sum of entropy of all neutrinos, shown in Figure~\ref{fig:total-S} increases with the systems size. We further compare the scaling of entanglement in two-flavor and three-flavor case under same initial state, {\it i.e.} all neutrinos initially in electron flavor $\ket{\nu_e}^{\otimes N}$. The difference between 2-flavor and 3-flavor case are increasing with an increase in the number of neutrinos in the ensemble signifying the importance of three-flavor calculations. As shown in Ref.~\cite{Patwardhan:2021rej}, in the two-flavor case, the difference between the survival probability of the highest frequency neutrino in the mean-field and many-body calculations increases with the number of neutrinos. This can be attributed to the highest frequency neutrino moving close to the center of the spectral split with increasing neutrino number. This motivated further exploration of the relationship between spectral splits and entanglement entropy for two flavors, and observing a similar trend in three flavors thereby serves as motivation for the current study.

To have a clear illustration of the number and position of spectral splits, we need to simulate a system of tens of neutrinos. However, due to the computational limitations of simulating such large Hilbert space (scaling as $3^N$ here, as opposed to $2^N$ in the two-flavor case), our many-body calculations are restricted to a system of seven neutrinos. Therefore, we rely on the mean-field calculations for a system of twelve neutrinos (Sec.~\ref{sec:SSMF}) to better interpret the spectral splits and later compare the results with the ones from many-body simulations for a smaller ensemble (Sec.~\ref{sec:SSMB}).

\subsection{Spectral Splits in the Mean-Field Calculations}
\label{sec:SSMF}

\begin{figure*}[t!]
    \centering
    \includegraphics[width=0.3\linewidth]{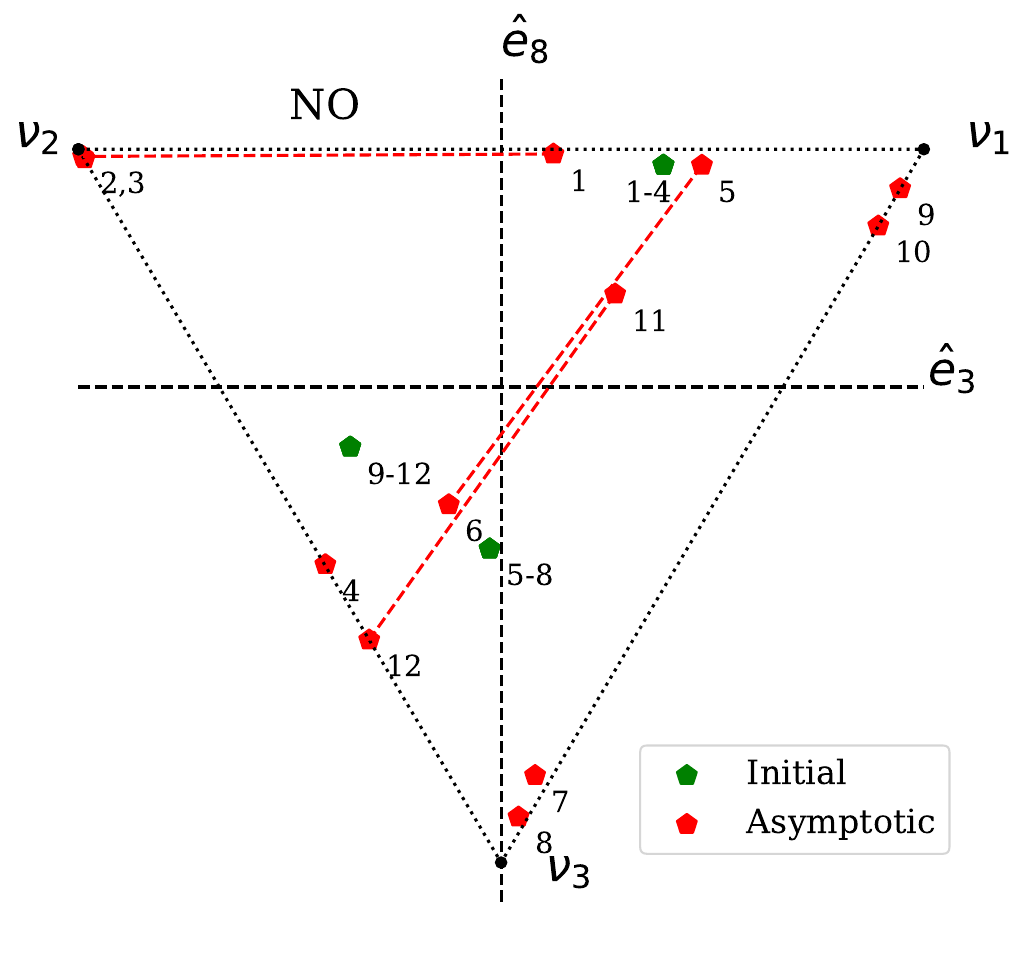}
    \includegraphics[width=0.3\linewidth]{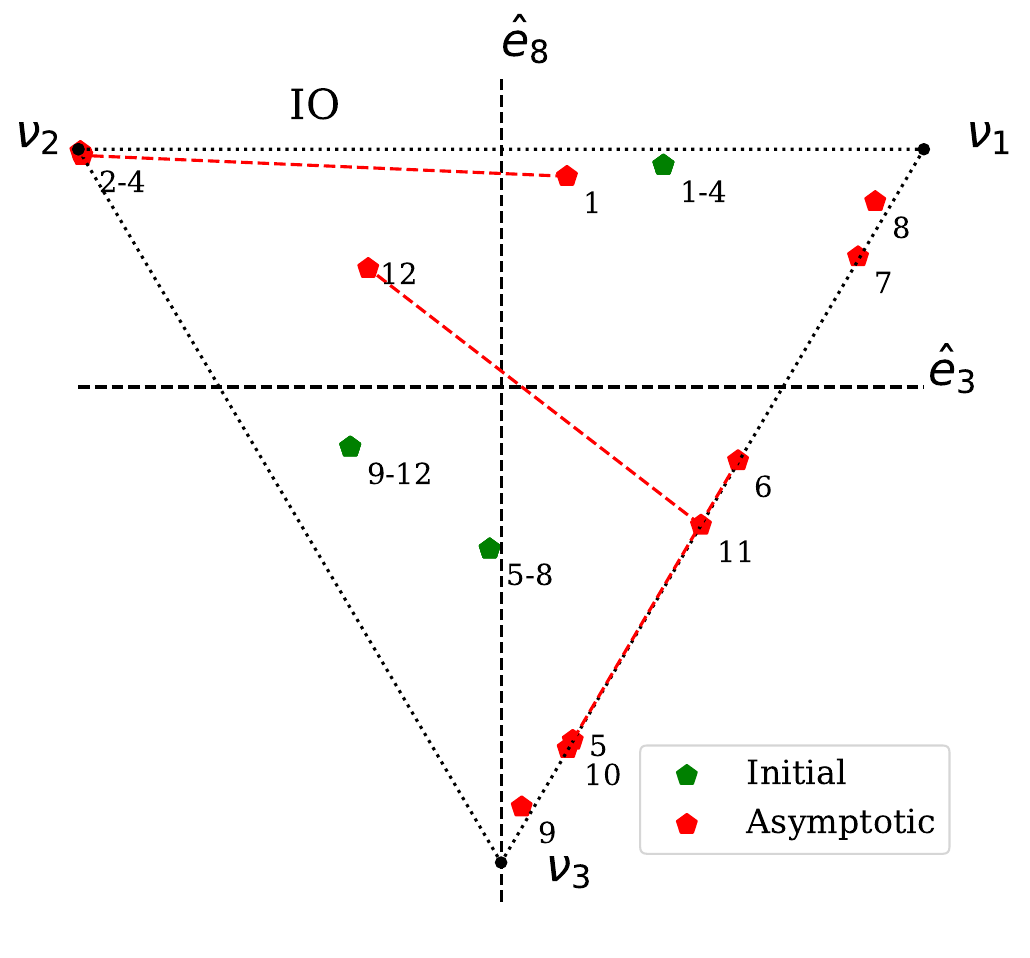}
    \includegraphics[width=0.3\linewidth]{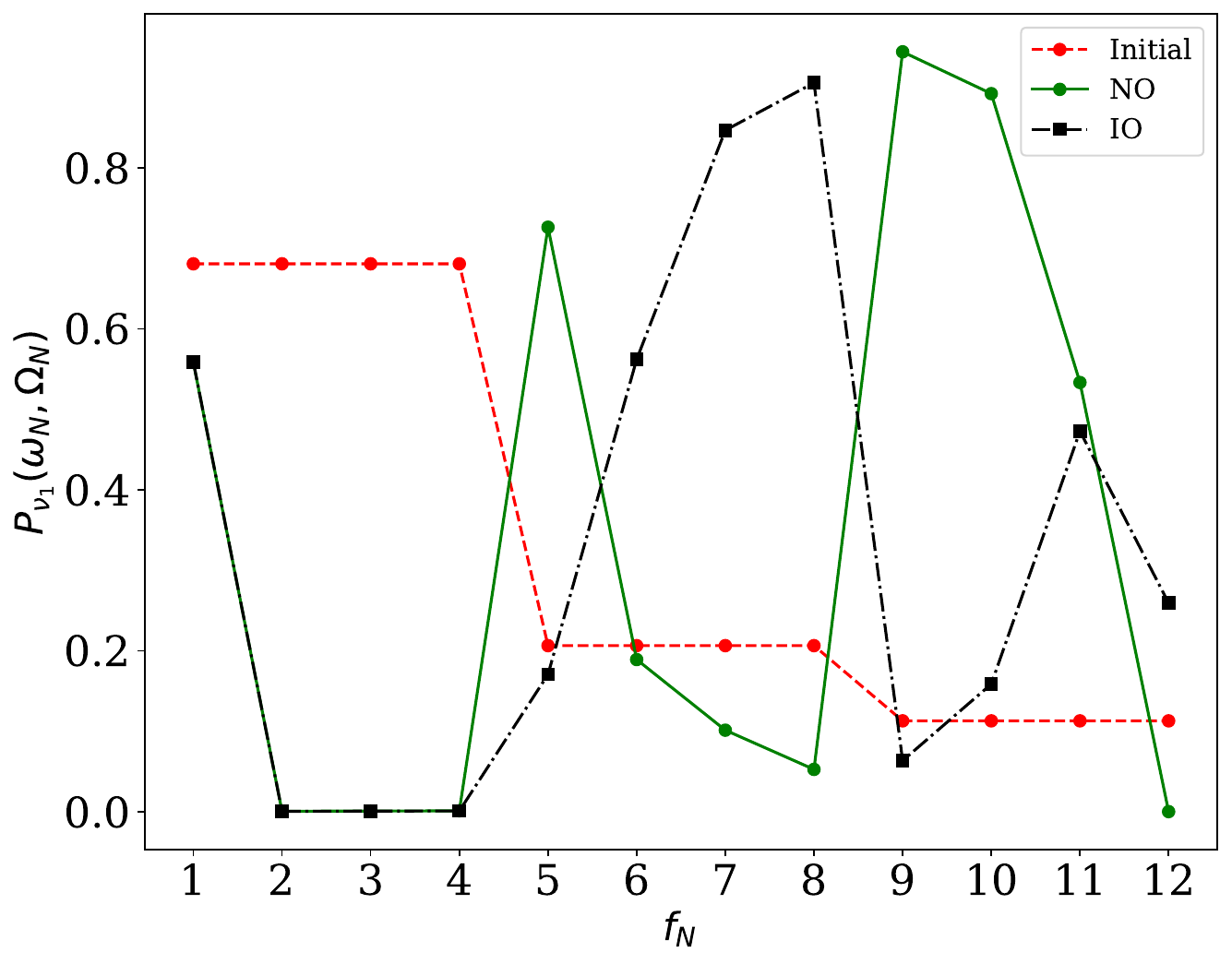}
    \caption{The asymptotic $P_3$ and $P_8$ values for $N=12$ neutrino system within the mean-field calculations with initial state $\ket{\nu_e^{\otimes 4}\nu_\mu^{\otimes 4}\nu_\tau^{\otimes 4}}$ for the NO (\emph{left panel}) and the IO (\emph{middle panel}) cases. \emph{Right panel:} The initial (red circles) and asymptotic values of $P_{\nu_1}$ for NO (green circles) and IO (black squares) as a function of the neutrino frequency mode number $f_N$.}
    \label{fig:N12_emt}
\end{figure*}

To visualize at what frequency neutrinos undergo a spectral split, we plot the asymptotic values of $P_3$ and $P_8$ in the $\hat{e}_3\text{--}\hat{e}_8$ plane. At the vertices of this triangle a neutrino is in one of the three mass eigenstates ($\nu_1,\nu_2$, and $\nu_3$) without admixture of other states. On the edge, a neutrino will have contribution of two mass eigenstates and zero contribution of the remaining third mass eigenstate. A spectral split emerges when some of the neutrinos from the ensemble move towards the first mass eigenstate $(\nu_1)$ and other ones towards $\nu_2$ or $\nu_3$. We begin with a system of 12 interacting neutrinos within the mean-field and initially all in electron flavor, {\it i.e.}, $\ket{\nu_e^{\otimes 12}}$. For such a system, the results are shown in Figure~\ref{fig:N12_e}. In the normal mass ordering (NO) case (left panel of the Fig.~\ref{fig:N12_e}), the neutrinos in the frequency modes $\omega_1$--$\omega_8$ lie at the vertex $\nu_1$. The remaining neutrinos with frequency modes $\omega_9$--$\omega_{12}$ are closer to $\nu_2$. Therefore, a single split occurs between neutrinos with frequency modes $\omega_8\leftrightarrow\omega_9$. Similar observation can be made in the IO (middle panel of the Fig.~\ref{fig:N12_e}) case except that the width of split is smaller as compared to NO case. 

Spectral splits can also be explained by looking at the probability of neutrinos in ensemble to be found in the first mass eigenstate $P_{\nu_1}$ in the asymptotic limit. As can be seen in the right panel of the Fig.~\ref{fig:N12_e}, the neutrinos split into two sectors $0.5< P_{\nu_1}\le 1$ and $0\le P_{\nu_1}<0.5$ at the frequencies $\omega_8\leftrightarrow\omega_9$ and hence explain the single spectral split. For an initial state with half of the neutrinos in electron flavor and half in muon flavor, i.e., $\ket{\nu_e^{\otimes 6}\nu_{\mu}^{\otimes 6}}$, two spectral splits emerge (not shown here). In the case of two-flavor approximation, the number of splits is the same for these two initial states, {\it i.e.}, one in the case of $\ket{\nu^{\otimes 12}}$ and two in the case of $\ket{\nu_e^{\otimes 6}\nu_{\mu}^{\otimes 6}}$~\cite{Patwardhan:2021rej}.

We further consider a more general initial state for an ensemble of 12 interacting neutrinos with 4 neutrinos in electron, muon and tau flavor each, {\it i.e.}, $\ket{\nu_e^{\otimes 4}\nu_{\mu}^{\otimes 4}\nu_{\tau}^{\otimes 4}}$. This initial state is not possible to study under two-flavor approximation and therefore is unique in the three-flavor case. The left and middle panels of Fig.~\ref{fig:N12_emt} show the asymptotic $P_3$ and $P_8$ values. In the case of NO and IO, three spectral splits emerge at $\omega_1\leftrightarrow\omega_2, \omega_5\leftrightarrow\omega_6$ and $\omega_{11}\leftrightarrow\omega_{12}$. Similar conclusions can be drawn from the $P_{\nu_1}$ shown in the right panel of Fig.~\ref{fig:N12_emt}. The split at $\omega_{11}\leftrightarrow\omega_{12}$ is much narrower in IO as compared to NO as can be seen from the $P_{\nu_1}$ values.

\begin{figure*}[t]
    \centering
    \includegraphics[width=0.9\columnwidth]{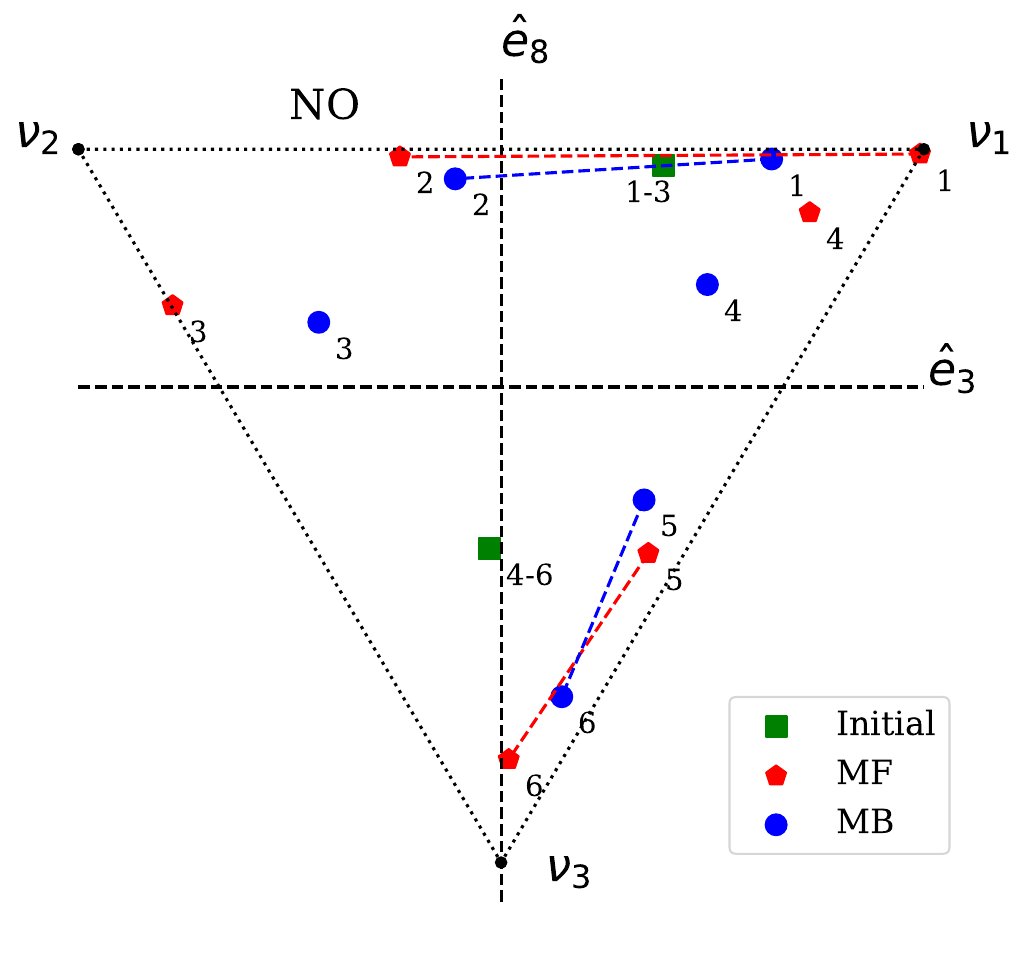}
    \includegraphics[width=0.9\columnwidth]{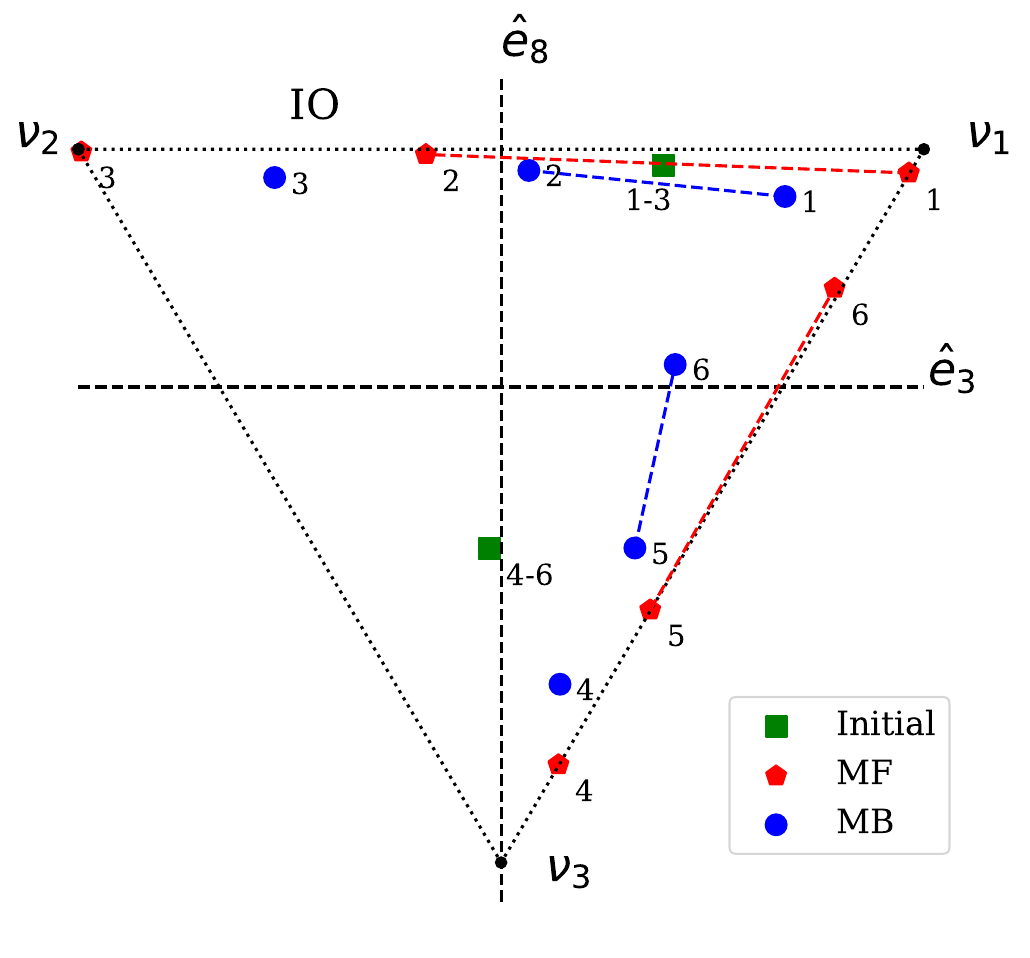}
    \caption{The asymptotic $P_{3}$ and $P_{8}$ values for $N=6$ neutrino system in the mean-field (MF) and many-body (MB) calculations with initial state $\ket{\nu_e^{\otimes 3}\nu_\mu^{\otimes 3}}$ for the NO (\emph{left panel}) and IO (\emph{right panel}). The spectral splits are represented by red and blue dashed lines in MF and MB, respectively. The initial states are shown by green squares.}
    \label{fig:N6_em}
\end{figure*}

\begin{figure*}
    \centering
    \includegraphics[width=0.9\columnwidth]{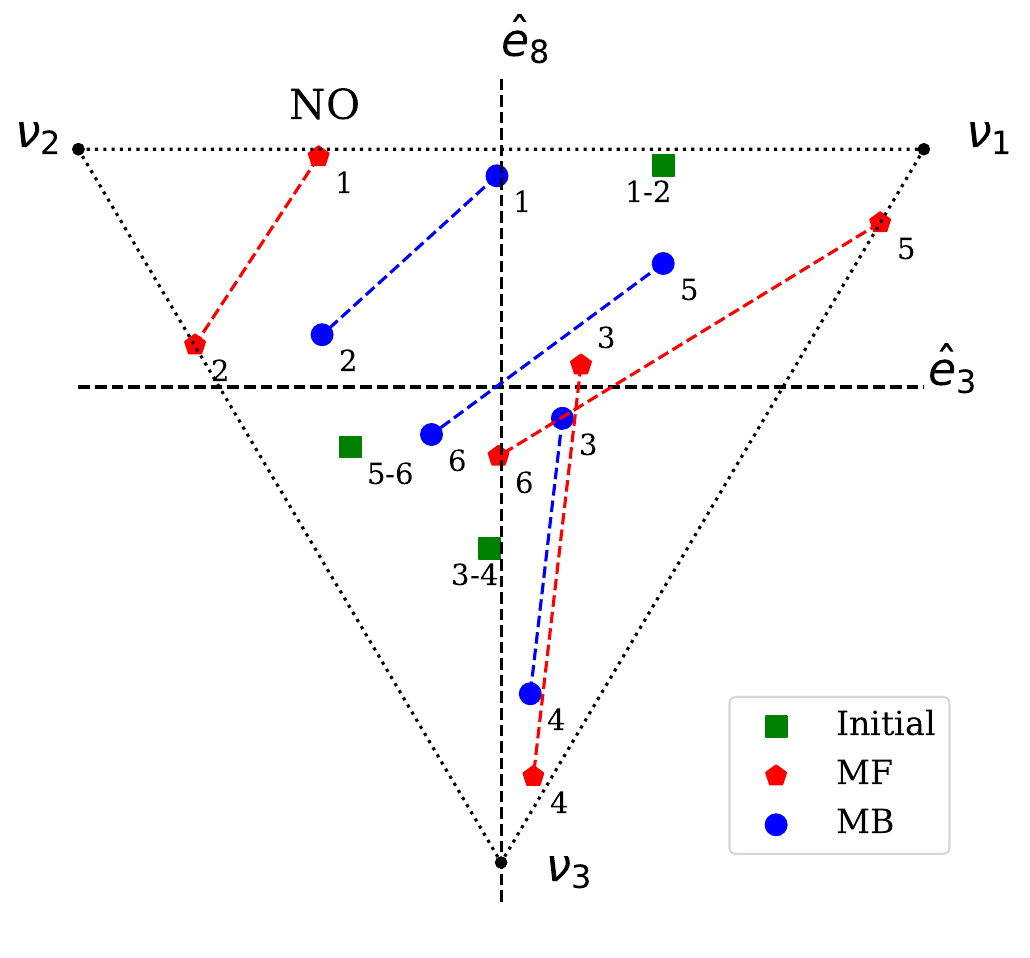}
    \includegraphics[width=0.9\columnwidth]{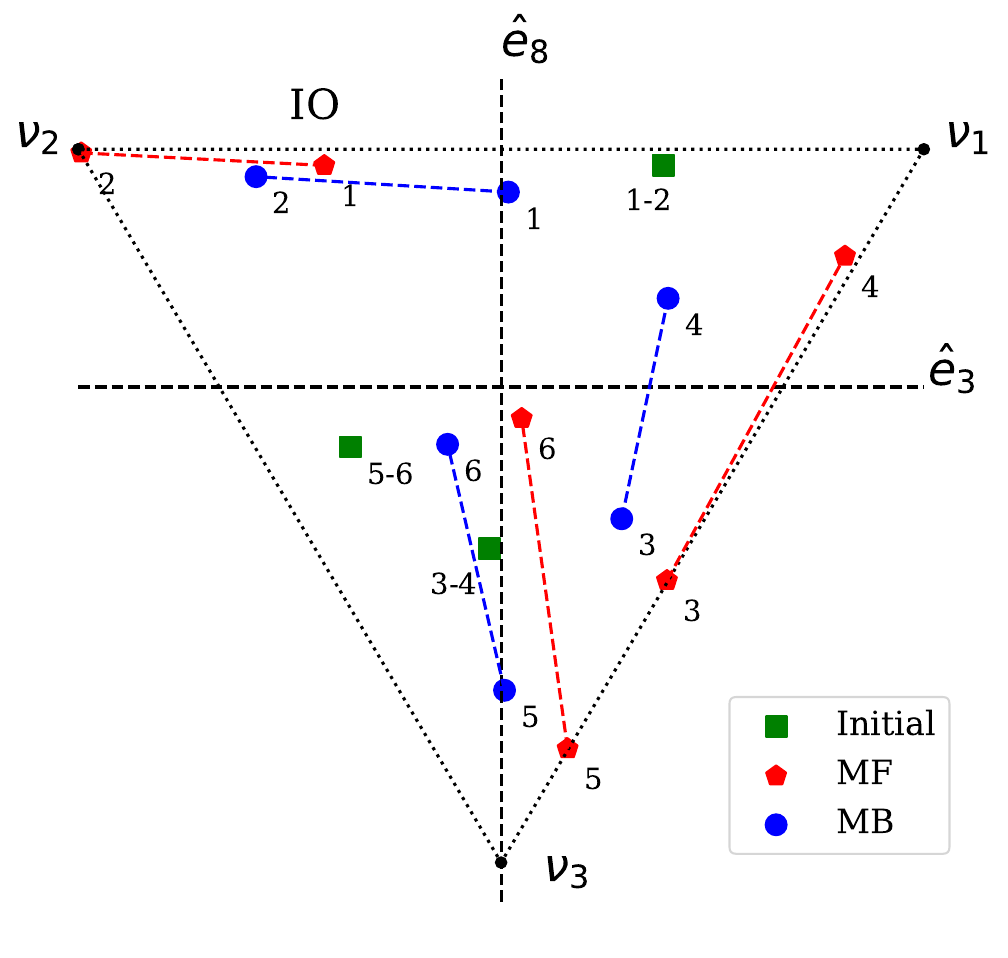}
    \caption{The asymptotic $P_3$ and $P_8$ values for $N=6$ neutrino system in the mean-field (MF) and many-body (MB) calculations with initial state $\ket{\nu_e^{\otimes 2}\nu_\mu^{\otimes 2}\nu_\tau^{\otimes 2}}$ for the NO (\emph{left panel}) and IO (\emph{right panel}). The spectral splits are represented by red and blue dashed lines in MF and MB, respectively. The initial states are shown by green squares.}
    \label{fig:N6_emt}
\end{figure*}

\subsection{Spectral Splits in the Many-Body Calculations}
\label{sec:SSMB}

The main goal of this work is to study the spectral splits in the many-body case and search for the correlations between spectral splits and entanglement entropy. We consider an ensemble of six interacting neutrinos since the spectral splits are more convenient to locate for an ensemble with the number of neutrinos as multiple of the number of flavors. We compare the results with the mean-field case for different initial states. For an initial state with three neutrinos in electron and muon flavor each $\ket{{\nu_e}^{\otimes 3}\nu_{\mu}^{\otimes 3}}$, the results are shown in Figure~\ref{fig:N6_em}. Two spectral splits at the frequency modes $\omega_1\leftrightarrow\omega_2$ and $\omega_5\leftrightarrow\omega_6$ emerge in both the cases NO (left panel of Fig.~\ref{fig:N6_em}) and IO (right panel of Fig.~\ref{fig:N6_em}). Furthermore, similar to the two-flavor case~\cite{Patwardhan:2021rej}, the strength of the spectral splits is significantly smaller in the case of many-body treatment as compared to the mean-field approximation. 

Figure~\ref{fig:N6_emt} illustrates the results in the case of a more general initial state, {\it i.e.}, $\ket{\nu_e^{\otimes 2}\nu_{\mu}^{\otimes 2}\nu_{\tau}^{\otimes 2}}$. Similarly to the mean-field calculations, three spectral splits emerge both in the NO and IO for an initial state with neutrinos in all three flavors. However, the splits at frequency modes $\omega_1\leftrightarrow\omega_2$, are not emerging as clearly as in the case of twelve neutrino case (see Figure~\ref{fig:N12_emt}) signifying the importance of simulating a larger ensemble of interacting neutrinos in the many-body calculations. The other two splits at frequency modes $\omega_3\leftrightarrow\omega_4$ and $\omega_5\leftrightarrow\omega_6$ again emphasize that the strength of spectral splits in the many-body treatment is smaller than in the mean-field calculations. Furthermore, in Refs.~\cite{Dasgupta:2008prd,Dasgupta:2009prl} the multiple spectral splits were observed under the mean-field approximation for a three-flavor case. As shown above, this argument of multiple spectral splits in three-flavor case still holds under the many-body effects.

\begin{figure*}
    \centering
    \includegraphics[width=0.45\linewidth]{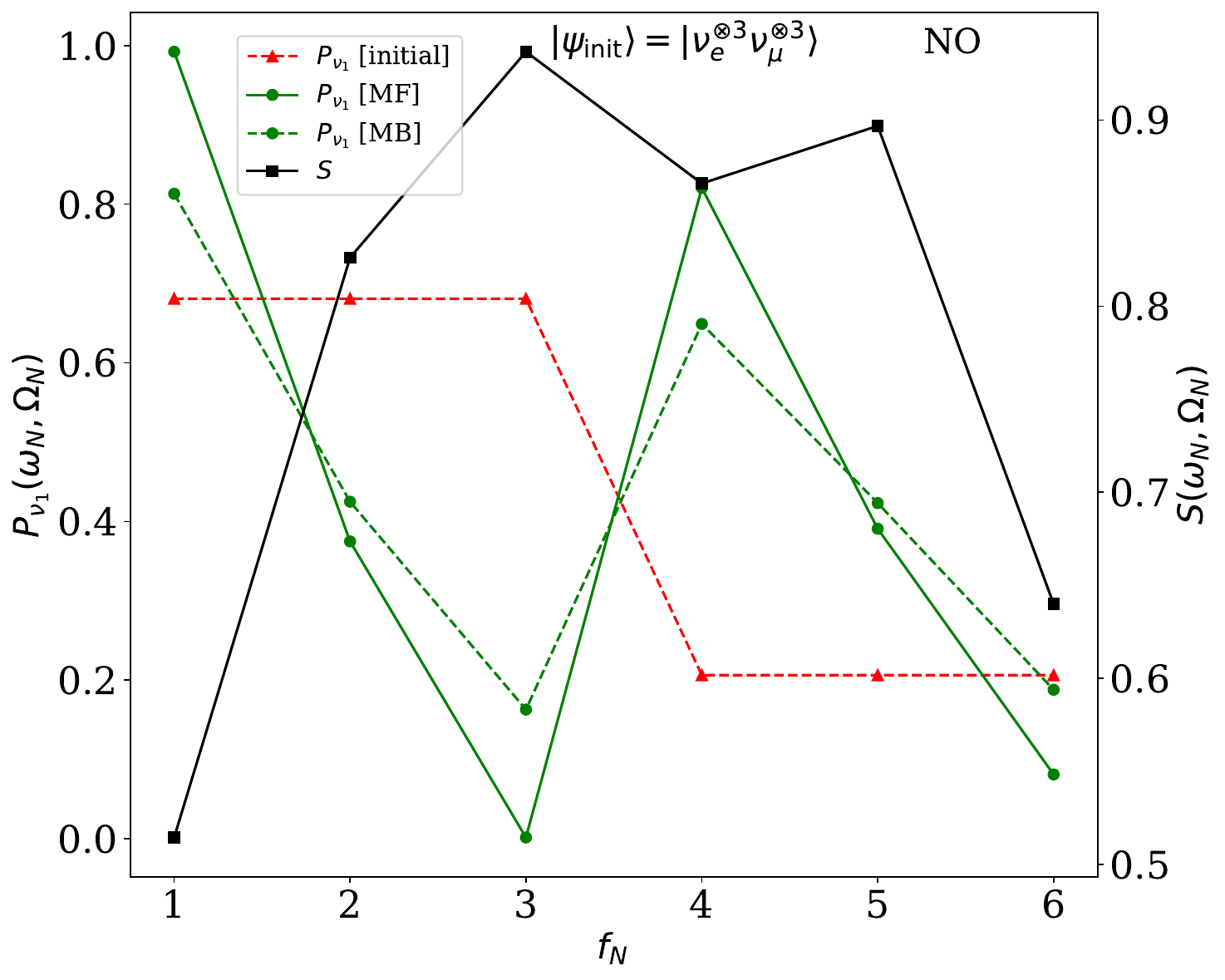}
    \includegraphics[width=0.45\linewidth]{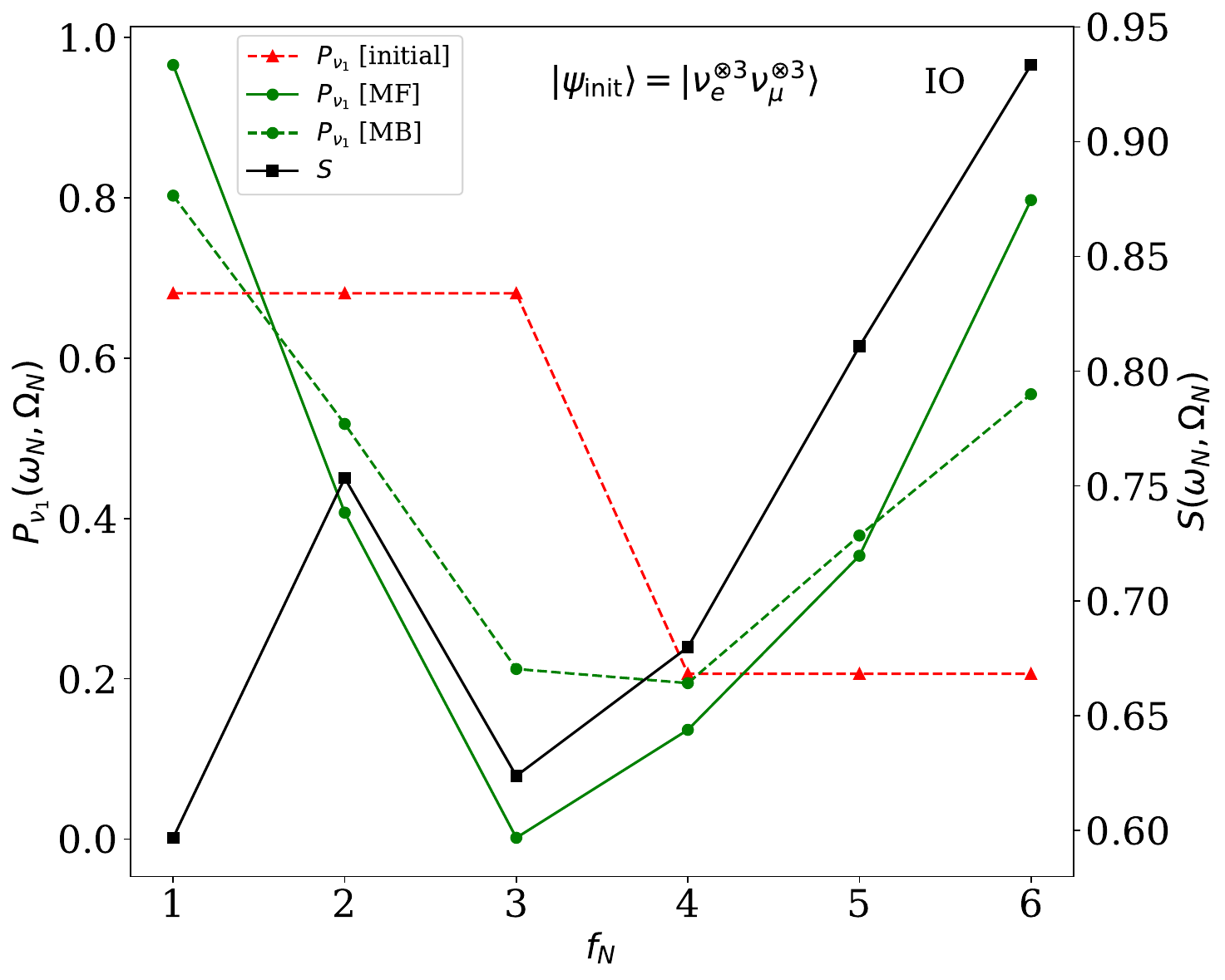}
    \caption{The asymptotic values of $P_3, P_8$ and entanglement entropy $S$ of neutrinos in the NO (left) and IO (right) cases for an ensemble of 6 interacting neutrinos with an initial state $\ket{\nu_e^{\otimes 3}\nu_{\mu}^{\otimes 3}}$. Here, $f_N=(\omega_N,\Omega_N)$ represents the neutrino in $N$th frequency mode.}
    \label{fig:N6_S13}
\end{figure*}


A very interesting correlation between the emergence of spectral splits and the entanglement was observed in the two-flavor collective neutrino oscillation simulations, that is, the neutrinos with the frequency at the spectral splits are the most entangled ones~\cite{Patwardhan:2021rej}. In this work, we study if that holds for a more realistic three-flavor case as well. 

Figures~\ref{fig:N6_S13} show the asymptotic values of entropy $(S)$ and $P_{\nu_1}$ for an ensemble of $6$ interacting neutrinos initially with neutrinos in two flavors $\ket{\nu_e^{\otimes3}\nu_{\mu}^{\otimes3}}$.
The most entangled neutrino lies closest to the centroid. In the case of NO (shown in Figure~\ref{fig:N6_S13} left panel), the spectral splits emerge at frequency modes $\omega_1\leftrightarrow\omega_2$ and $\omega_5\leftrightarrow\omega_6$ (see also Figure~\ref{fig:N6_em}), however, the neutrino with third frequency mode the the largest entanglement entropy. The neutrinos with $N=1$ and $N=6$, which lie at the spectral splits have very small entropy. For the same initial state but in the IO case (right panel of Fig.~\ref{fig:N6_S13}), the neutrino with $N=6$ is the most entangled one, but $N=1$ which lies at the spectral split has the minimum entanglement entropy. Furthermore, it was found in the two-flavor case that the deviation between the mean-field and many-body $P_{\nu_{1}}$ value is maximum for the neutrinos at the spectral splits and the most entangled ones. This does not hold true in the present three-flavor case. Similarly in the case of initial state $\ket{\nu_e^{\otimes2}\nu_{\mu}^{\otimes2}\nu_{\tau}^{\otimes2}}$, where all 6 neutrinos lie at the spectral splits (see Figure~\ref{fig:N6_emt}), we make similar observations (results not shown here). Therefore, the conclusions from the two-flavor case, {\it i.e.}, the neutrinos at spectral splits are the most entangled ones does not seem to be necessarily holding true in the more realistic three-flavor treatment, particularly if one only considers the behavior of $P_{\nu_1}$.

There can be several possible reasons behind the disagreement between the most entangled neutrino and spectral split frequencies. We suspect that the interpretation of spectral splits in $\nu_1$ vs $\nu_2$--$\nu_3$ sector is not a full picture to draw conclusions about the entanglement. A more detailed analysis, involving overlap probabilities of all three mass eigenstates, $P_{\nu_1}, P_{\nu_2}$, and $P_{\nu_3}$, would be necessary to uncover a correlation between the spectral split locations and the entanglement properties of neutrinos. A second likely reason is that we quantify the quantum correlations only in terms of the bipartite entanglement. Multipartite entanglement might provide better insights into its correlations with spectral splits. All these factors should be investigated in detail.


\section{Conclusions}
\label{sec:Conclusions}
We study the spectral splits in three-flavor collective neutrino oscillations within the many-body picture for the first time, performing simulations of ensembles of up to seven interacting neutrinos.
The total entanglement entropy increases with the size of the ensemble and deviations from the two-flavor case also increase. Multiple spectral splits are observed in the three-flavor many-body case agreeing with the mean-field results. The spectral splits are located at the same frequencies both in the mean-field and many-body case; however, the strength of the splits is smaller in the many-body case as compared to the mean-field case. 
Furthermore, we analyze the results in both normal and inverted mass orderings and find that the spectral splits emerge at the same frequency in both mass orderings.
Unlike in the two-flavor many-body calculations~\cite{Patwardhan:2021rej}, we do not find that the neutrinos with the frequency modes at the spectral splits are necessarily the most entangled ones in the system. That finding holds regardless of if the initial state is consisting of two or three neutrino flavors in the three-flavor case.  However, a concurrent analysis of splits not just between $\nu_1$ and the other states, but also in orthogonal directions (e.g., $\nu_2$--$\nu_3$ splits) would required to draw definitive conclusions in this regard.

 Our findings open new avenues for exploring different entanglement measures in collective neutrino oscillations and other quantum many-body problems in general. It would be interesting to see if such correlations exist in other phenomena and how they behave with different number of energy levels in the system. Such studies will help us better understand the role of entanglement in quantum systems.

\begin{acknowledgments}
Support for this work was partly provided through Scientific Discovery through Advanced Computing (SciDAC) program funded by U.S. Department of Energy, Office of Science, Advanced Scientific Computing Research and Nuclear Physics. It was partly performed under the auspices of the U.S. Department of Energy by the Lawrence Livermore National Laboratory under Contract No. DE-AC52-07NA27344. Computing support for this work came from the Lawrence Livermore National Laboratory (LLNL) Institutional Computing Grand Challenge program. This work was supported in part by the U.S.~Department of Energy, Office of Science, Office of High Energy Physics, under Award No.~DE-SC0019465 and in part by the National Science Foundation Grants No. PHY-1806368, PHY-2020275, PHY-2108339 and PHY-2411495.
The work was also partially  supported by the Neutrino Theory Network Program Grant No.~DE-AC02-07CHI11359. The work of AVP was supported by the U.S. Department of Energy (DOE) under grant DE-FG02-87ER40328 at the University of Minnesota.
\end{acknowledgments}


\appendix

\section{Mean-Field equations}\label{sec:mf_Eq}
We expand the Eq.~\eqref{eq:mf_final} in terms of all 8 components of polarization vector $P$. Using the following values of the structure coefficients
\begin{eqnarray}
    f^{123} &=& 1,\nonumber\\
    f^{147} &=& - f^{156} = f^{246} = f^{257} = f^{345} = - f^{367} = \frac{1}{2},\nonumber\\
    f^{458} &=& f^{678} = \frac{\sqrt{3}}{2}.
\end{eqnarray}
we get
\begin{widetext}
\begin{eqnarray}
    \frac{\partial P_{1}}{\partial t}&=&(B_{2}P_{3}-B_{3}P_{2}+\frac{1}{2}\left[B_{4}P_{7}-B_{7}P_{4}-B_{5}P_{6}+B_{6}P_{5}\right])+\mu(r)\left(\Pi_{2}P_{3}-\Pi_{3}P_{2}+\frac{1}{2}\left[\Pi_{4}P_{7}-\Pi_{7}P_{4}-\Pi_{5}P_{6}+\Pi_{6}P_{5}\right]\right)\nonumber\\
    \frac{\partial P_{2}}{\partial t}&=&(B_{3}P_{1}-B_{1}P_{3}+\frac{1}{2}\left[B_{4}P_{6}-B_{6}P_{4}+B_{5}P_{7}-B_{7}P_{5}\right])+\mu(r)\left(\Pi_{3}P_{1}-\Pi_{1}P_{3}+\frac{1}{2}\left[\Pi_{4}P_{6}-\Pi_{6}P_{4}+\Pi_{5}P_{7}-\Pi_{7}P_{5}\right]\right)\nonumber\\
    \frac{\partial P_{3}}{\partial t}&=&(B_{1}P_{2}-B_{2}P_{1}+\frac{1}{2}\left[B_{4}P_{5}-B_{5}P_{4}-B_{6}P_{7}+B_{7}P_{6}\right])+\mu(r)\left(\Pi_{1}P_{2}-\Pi_{2}P_{1}+\frac{1}{2}\left[\Pi_{4}P_{5}-\Pi_{5}P_{4}-\Pi_{6}P_{7}+\Pi_{7}P_{6}\right]\right)\nonumber\\
    \frac{\partial P_{4}}{\partial t}&=&\frac{1}{2}\left[B_{7}P_{1}-B_{1}P_{7}-B_{2}P_{6}+B_{6}P_{2}-B_{3}P_{5}+B_{5}P_{3}\right]+\frac{\sqrt{3}}{2}\left[B_{5}P_{8}-B_{8}P_{5}\right]\nonumber\\
    &~&+\mu(r)\left(\frac{1}{2}\left[\Pi_{7}P_{1}-\Pi_{1}P_{7}-\Pi_{2}P_{6}+\Pi_{6}P_{2}-P^{T}_{3}P_{5}+\Pi_{5}P_{3}\right]+\frac{\sqrt{3}}{2}\left[\Pi_{5}P_{8}-\Pi_{8}P_{5}\right])\right)\nonumber\\
\frac{\partial P_{5}}{\partial t}&=&\frac{1}{2}\left[B_{1}P_{6}-B_{6}P_{1}-B_{2}P_{7}+B_{7}P_{2}+B_{3}P_{4}-B_{4}P_{3}\right]+\frac{\sqrt{3}}{2}\left[B_{8}P_{4}-B_{4}P_{8}\right])\nonumber\\
&~&+\mu(r)\left(\frac{1}{2}\left[\Pi_{1}P_{6}-\Pi_{6}P_{1}-\Pi_{2}P_{7}+\Pi_{7}P_{2}+\Pi_{3}P_{4}-\Pi_{4}P_{3}\right]+\frac{\sqrt{3}}{2}\left[\Pi_{8}P_{4}-\Pi_{4}P_{8}\right]\right)\nonumber\\
\frac{\partial P_{6}}{\partial t}&=&\frac{1}{2}\left[B_{5}P_{1}-B_{1}P_{5}+B_{2}P_{4}-B_{4}P_{2}+B_{3}P_{7}-B_{7}P_{3}\right]+\frac{\sqrt{3}}{2}\left[B_{7}P_{8}-B_{8}P_{7}\right]\nonumber\\
&~&+\mu(r)\left(\frac{1}{2}\left[\Pi_{5}P_{1}-\Pi_{1}P_{5}+\Pi_{2}P_{4}-\Pi_{4}P_{2}+\Pi_{3}P_{7}-\Pi_{7}P_{3}\right]+\frac{\sqrt{3}}{2}\left[\Pi_{7}P_{8}-\Pi_{8}P_{7}\right]\right)\nonumber\\
\frac{\partial P_{7}}{\partial t}&=&\frac{1}{2}\left[B_{1}P_{4}-B_{4}P_{1}+B_{2}P_{5}-B_{5}P_{2}-B_{3}P_{6}+B_{6}P_{3}\right]+\frac{\sqrt{3}}{2}\left[B_{8}P_{6}-B_{6}P_{8}\right]\nonumber\\
&~&+\mu(r)\left(\frac{1}{2}\left[\Pi_{1}P_{4}-\Pi_{4}P_{1}+\Pi_{2}P_{5}-\Pi_{5}P_{2}-\Pi_{3}P_{6}+\Pi_{6}P_{3}\right]+\frac{\sqrt{3}}{2}\left[\Pi_{8}P_{6}-\Pi_{6}P_{8}\right]\right)\nonumber\\
\frac{\partial P_{8}}{\partial t}&=&\frac{\sqrt{3}}{2}\left[B_{4}P_{5}-B_{5}P_{4}+B_{6}P_{7}-B_{7}P_{6}\right]+\mu(r)\left(\frac{\sqrt{3}}{2}\left[\Pi_{4}P_{5}-\Pi_{5}P_{4}+\Pi_{6}P_{7}-\Pi_{7}P_{6}\right]\right)
\end{eqnarray}
\end{widetext}

Here, $B_i\equiv B(\omega_i,\Omega_i)$ and $P(i)\equiv P_i$.

\bibliography{3-level}

\end{document}